\documentclass[aps,prl,twocolumn,showpacs,amsmath,amssymb]{revtex4-1}
\usepackage[utf8]{inputenc}
\usepackage{amsmath}
\usepackage{graphicx}
\usepackage{hyperref}
\usepackage{bbold}
\usepackage{braket}
\usepackage{etoolbox}
\usepackage{cancel}
\usepackage{mathtools,color}
\newcommand{\x}{{\rm x}}
\newcommand{\y}{{\rm y}}
\newcommand{\z}{{\rm z}}
\newcommand{\dd}{{\rm d}}
\newcommand{\tr}{{\rm tr}}
\newcommand{\op}[1]{{#1}}
\newcommand{\s}{\op{s}}

\begin{document}
    
    \title{Kardar-Parisi-Zhang physics in the quantum Heisenberg magnet}
    \author{Marko Ljubotina, Marko Žnidarič, and Tomaž Prosen}
    \affiliation{Physics Department, Faculty of Mathematics and Physics, University of Ljubljana, 1000 Ljubljana, Slovenia}
    
    \begin{abstract}
      Equilibrium spatio-temporal correlation functions are central to understanding weak nonequilibrium physics. In certain local one-dimensional classical systems with three conservation laws they show universal features. Namely, fluctuations around ballistically propagating sound modes can be described by the celebrated Kardar-Parisi-Zhang (KPZ) universality class. Can such a universality class be found also in quantum systems? By unambiguously demonstrating that the KPZ scaling function describes magnetization dynamics in the SU(2) symmetric Heisenberg spin chain we show, for the first time, that this is so. We achieve that by introducing new theoretical and numerical tools, and make a puzzling observation that the conservation of energy does not seem to matter for the KPZ physics.
    \end{abstract}
    \maketitle

    \textbf{Introduction.--} 
    Universality -- where different systems can be described by the same underlying mathematical structure -- is at the core of our understanding of nature. 
    For instance, the properties of any thermalizing system can be described by the same equilibrium ensembles of statistical physics. 
    Out of equilibrium less is known in general, in a way justifiably so, simply because the world of nonequilibrium is much richer. 
    One of the more famous universality classes that can (among other) describe various nonequilibrium phenomena~\cite{livibook} is that of the Kardar-Parisi-Zhang (KPZ) equation. 
    The KPZ equation was originally introduced to describe stochastic growth of surfaces~\cite{kardar86}, and is a diffusion equation with the simplest possible nonlinearity (relevant at large scales) and an additional white noise term (equivalently, the surface's slope is described by the stochastic Burgers equation). 
    Besides describing surface dynamics it can be found in various contexts, ranging from exclusion processes to random matrix theory, for review see~\cite{kpzrev1}. 
    The KPZ equation itself harbors rich mathematical problems~\cite{kpzrev2}.
    
    Nonequilibrium physics is one of the more propulsive areas of today's theoretical physics. 
    Close to equilibrium one can use Green-Kubo formulas and express nonequilibrium properties in terms of equilibrium correlation functions~\cite{kubobook}. 
    A downside to such an approach is that the calculation of spatio-temporal correlation functions is often very complicated. 
    Any possible universality in their long-time behavior would therefore be highly appreciated. 
    For classical fluids in one dimension such a picture has in fact been put forward~\cite{beijeren12,spohn14} in a form of nonlinear fluctuating hydrodynamics~\cite{spohnrev} that describes (anomalous) fluctuations around sound peaks due to nonlinearity in one-dimensional systems that have 3 conservation laws (momentum, energy and mass), and are in general nonintegrable. That fluctuations are indeed described by the KPZ scaling function~\cite{spohn04} has been verified in a number of classical systems~\cite{kulkarni13,mendl14,das14,das19,unpub}. So-far there has been no observation of the KPZ universality class scaling function in quantum systems.
    
    In this Letter we observe the KPZ scaling functions in an integrable quantum model that does not have any ballistic component. Namely, we show with an unprecedented accuracy (an order of magnitude larger than in simulations of classical systems) that an infinite temperature spin-spin correlation function in a paradigmatic SU(2) symmetric quantum Heisenberg chain has a KPZ form. Such accuracy is a result of two novelties: (i) using linear response formulation we show that one can calculate the equilibrium correlation function as an expectation value in a particular nonequilibrium state whose time evolution is easier to calculate, (ii) we directly treat an ensemble evolution, avoiding statistical averaging (as done in classical simulations), and which is, even more importantly, structurally stable. 
    In addition, to discern the role played by conserved quantities, we show that in an integrable trotterized Floquet generalization~\cite{vanicat18} of the model, that does not conserve the energy, the same KPZ scaling is observed. 
    We note that the KPZ scaling exponents have been observed in various stochastic quantum settings, like random quantum circuits~\cite{nahum17,nahum18} or noisy evolution~\cite{lamacraft18}.
    
    \textbf{The model.--}
    In classical systems the KPZ scaling function describes fluctuations around a sound mode, whose width scales as $\sim t^{1/z}$ with a dynamical exponent $z=\frac{3}{2}$. Therefore, to observe it one has to move to a ballistically moving reference frame, which, if the velocity is not known analytically, can introduce numerical inaccuracies. We are therefore going to look for KPZ physics at infinite temperature in the one-dimensional Heisenberg spin-$\frac{1}{2}$ chain at zero magnetization (half-filling) where the ballistic contribution is zero due to the spin-flip (particle-hole) symmetry and the spin transport shows a KPZ dynamical scaling exponent $z=\frac{3}{2}$. Namely, such superdiffusive magnetization transport has been observed in a nonequilibrium steady state where the current scales as $j \sim 1/L^{z-1}$~\cite{prl11} as well as in the spreading of an inhomogeneous initial state where the width scales as $\sigma \sim t^{1/z}$~\cite{ljubotina17}. The local Hamiltonian density is
    \begin{equation}
        \op{h}_{r,r+1}=J\left(\s^\x_{r}\s^\x_{r+1}+\s^\y_{r}\s^\y_{r+1}+\s^\z_{r}\s^\z_{r+1}\right)\,,
        \label{eq:localint}
    \end{equation}
    where $s^{\alpha}_r=\frac{1}{2}\sigma^{\alpha}_r$, $\alpha\in\{\x,\y,\z\}$, are spin operators (Pauli matrices) at site $r\in\{-\frac{L}{2},\ldots,\frac{L}{2}-1\}$. Theoretical explanation of the scaling exponent $z=\frac{3}{2}$ is still lacking,
    but consistent derivations within assumptions of generalized hydrodynamics were recently given~\cite{vasseur18}. In particular, it is possible to estimate the diffusion constant \cite{markomprl,denardis18,vasseur18} and prove its divergence, i.e. $z < 2$ \cite{enejprl}.
    
    Here, in order to observe precise spatio-temporal profiles of spin and current densities, we will consider two dynamical setings:
    {\em continuous} time evolution
    $U^t=e^{-i Ht}$ generated by 
    $H=\sum_{r=-L/2}^{L/2-2} h_{r,r+1}$ (where we set $J=1$) or {\em discrete} time evolution with one step propagator
    $\op{U}=\op{U}_{\rm e}\op{U}_{\rm o}$, with $\op{U}_{\rm o}=e^{-i\sum_{r}\op{h}_{2r-1,2r}}$ and $\op{U}_{\rm e}=e^{-i\sum_{r}\op{h}_{2r,2r+1}}$ (where we use $J=\frac{\pi}{2}$, and where one also observes the superdiffusive
    scaling $z=\frac{3}{2}$~\cite{ljubotina19}).
    Both settings are characterized by both a global $SU(2)$ symmetry and integrability. 
    
    In order to study transport we must derive the expressions for the local spin current density operators for both the continuous-time and discrete-time models. 
    The former is the standard spin current in the Heisenberg model $\op{j}_r=\s_r^\x\s_{r+1}^\y-\s_r^\y\s_{r+1}^\x$
    which fulfills the continuity equation $\frac{\dd \s^\z_r}{\dd t}=\op{j}_{r-1}-\op{j}_{r}$. 
    The current in the discrete-time model turns out to be slightly more complicated, with the operator being different on odd and even sites due to the staggered nature of the propagator $U$. 
    The two currents densities satisfy a pair of continuity equations 
    \begin{equation}
        \begin{aligned}
            &\op{U}^\dagger\op{M}_{2r}\op{U}-\op{M}_{2r}=\op{j}_{2r-1}^{\rm o}-\op{j}_{2r+1}^{\rm o}\,,\\
            &\op{U}^\dagger\op{M}_{2r-1}\op{U}-\op{M}_{2r-1}=\op{j}_{2r-2}^{\rm e}-\op{j}_{2r}^{\rm e}\,,
        \end{aligned}
        \label{eq:disc_continuity}
    \end{equation}
    where $\op{M}_r=\s_{r}^\z+\s_{r+1}^\z$.
    The simpler odd current can then be seen to take the form
    \begin{equation}
        \begin{aligned}
            \op{j}_{2r-1}^{\rm o}=2\sin(J) j_{2r-1} -\frac{1}{2}\sin^2(J/2)(\s^\z_{2r}-\s^\z_{2r-1})\,,
        \end{aligned}
        \label{eq:disc_curr}
    \end{equation}
    whereas the even current is simply the odd current propagated by half a time step $\op{j}_{2r}^{\rm e}=\op{U}^\dagger_{\rm e} \op{j}_{2r}^{\rm o}\op{U}_{\rm e}$ and acts on 4 adjacent sites. 
    
    \begin{figure}[ht!]
        \centering
        \includegraphics[width=0.98\linewidth]{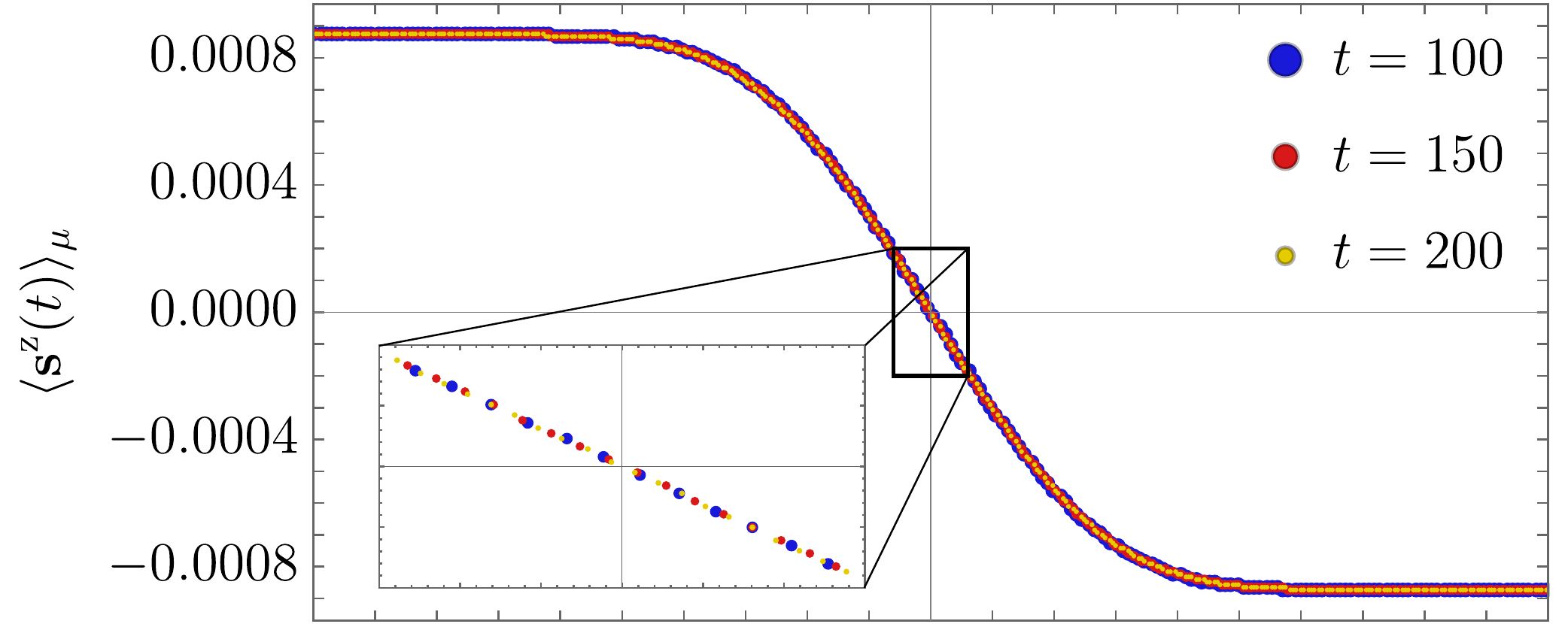}
        \includegraphics[width=0.98\linewidth]{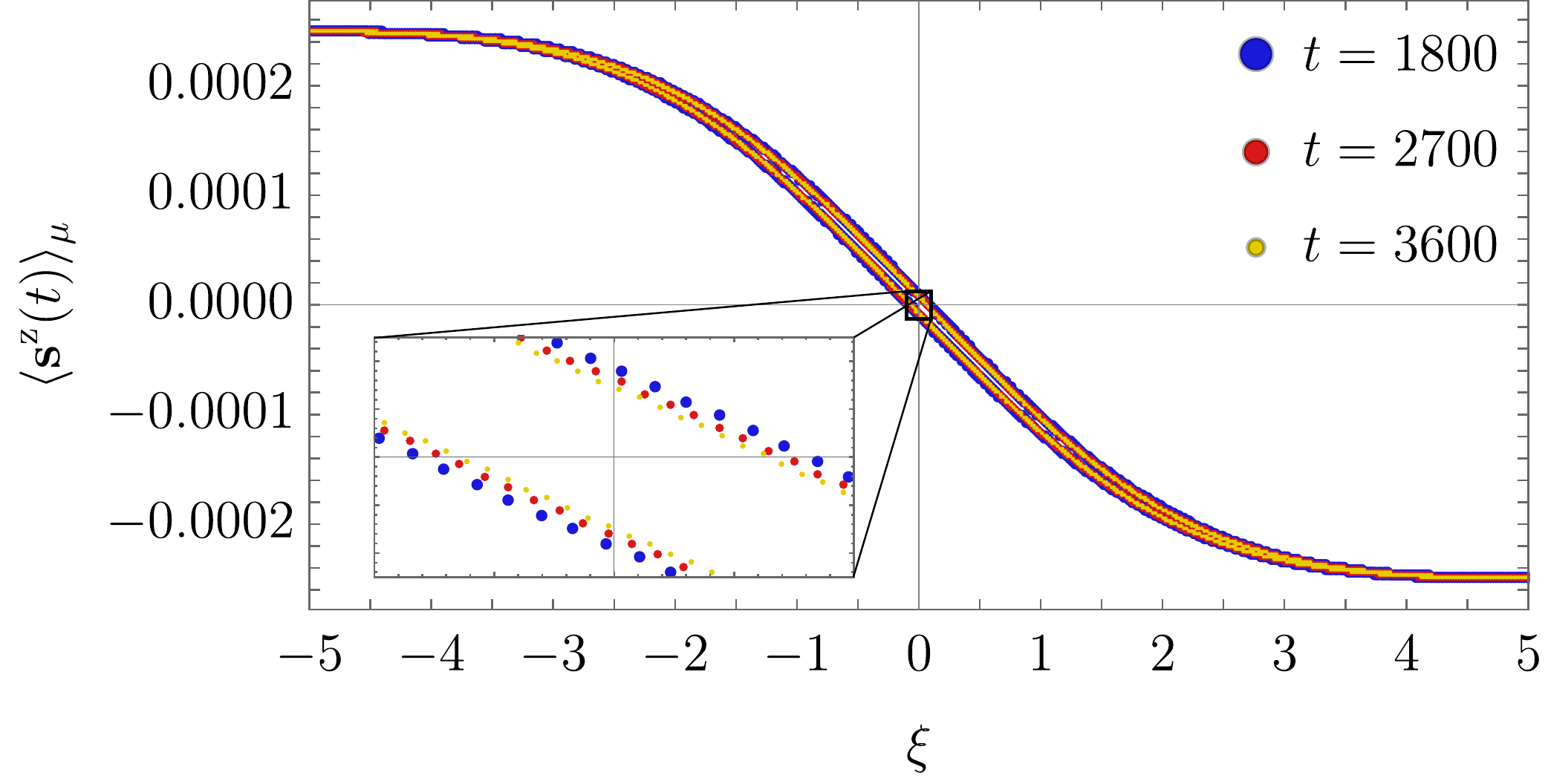}
        \caption{\noindent
            Collapse of spin profiles for the continuous-time (top) and discrete-time (bottom) model in terms of the scaling parameter $\xi=x/t^{2/3}$ shown for several times. 
            The continuous-time simulation was performed on a spin chain of length $L=400$ with bond dimension $\chi=400$ and polarization $\mu=0.0017$.
            The discrete-time simulation was performed with $L=7200$, $\chi=256$ and $\mu=0.0005$. The same parameters are used in other figures. In the discrete case there is an additional Floquet even-odd splitting whose size decays as $t^{-1/3}$ (the inset). 
        }
        \label{fig:profiles}
    \end{figure}
    
    We begin by preparing our system in a weakly polarized domain-wall mixed initial state~\cite{ljubotina17}
    \begin{equation}
        \begin{aligned}            
            \rho(t=0)&\propto \rho_\mu=\left(e^{\mu\s^\z}\right)^{\otimes L/2}\otimes\left(e^{-\mu\s^\z}\right)^{\otimes L/2}.
        \end{aligned}
        \label{eq:init_state}
    \end{equation}
    An example of time evolution for both models is shown in Fig.~\ref{fig:profiles}, using the scaling variable $\xi=\frac{r}{t^{1/z}}, z=\frac{3}{2}$. 
    While this choice of the initial state provides a numerically stable and efficient way to study spin transport~\cite{ljubotina17}, we emphasize that for our purposes it provides us with an efficient way to study the infinite-temperature spin-spin correlation function $\langle\s^\z_0\s^\z_r(t)\rangle$, where $A(t)\equiv U^{-t}AU^t$ and $\langle\,\boldsymbol{\cdot}\,\rangle\equiv 2^{-L}\tr(\cdot)$ denotes the infinite-temperature expectation value. 
    We explain that in the following section.
        
    \textbf{Linear response.--}
    We start by expanding the initial state (\ref{eq:init_state}) to linear order in $\mu$, evolving it in time, and writing down the expectation value for a single spin,
    \begin{equation}
        \langle\s_r^\z(t)\rangle_\mu=-\mu\sum_{r'} \theta_{r'} \langle\s_r^\z(t) \s_{r'}^\z\rangle+\mathcal{O}(\mu^2)\,,
        \label{eq:lin1}
    \end{equation}
    where we introduced $\langle\,\boldsymbol{\cdot}\,\rangle_\mu = \tr[(\cdot)\rho_\mu]/\tr\rho_\mu$ as the expectation value in the weak domain-wall initial state (\ref{eq:init_state}) and $\theta_{r}\equiv 1 (-1)$ for $r\ge0 (<0)$. 
    Accounting for the translational invariance of the infinite-temperature expectation value we obtain
    \begin{eqnarray}
        &\langle\s_{r-1}^\z(t)\rangle_\mu-\langle\s_{r}^\z(t)\rangle_\mu\approx
        \mu\langle\s^\z_r(t)\sum_{r'}\theta_{r'}(\s^\z_{r'}-\s^\z_{r'+1})\rangle\nonumber \\
        &=2\mu \langle\s^\z_r(t)\s^\z_0\rangle - 2\mu \langle\s^\z_r(t)\s^\z_{-L/2}\rangle.
        \label{eq:lin2}
    \end{eqnarray}
    In the thermodynamic limit $L \to \infty$ the second term vanishes as there are no correlations across infinite distances, and using the cyclic property of the trace we get
    \begin{equation}
              \langle\s_0^\z(0)\s_r^\z(t)\rangle=\lim_{\mu\to0}\frac{\langle\s_{r-1}^\z(t)\rangle_\mu-\langle\s_{r}^\z(t)\rangle_\mu}{2\mu}\,.
        \label{eq:lin3}
    \end{equation}
    This is our first main result. 
    \begin{figure*}[t!]
        \centering
        \includegraphics[width=0.49\linewidth]{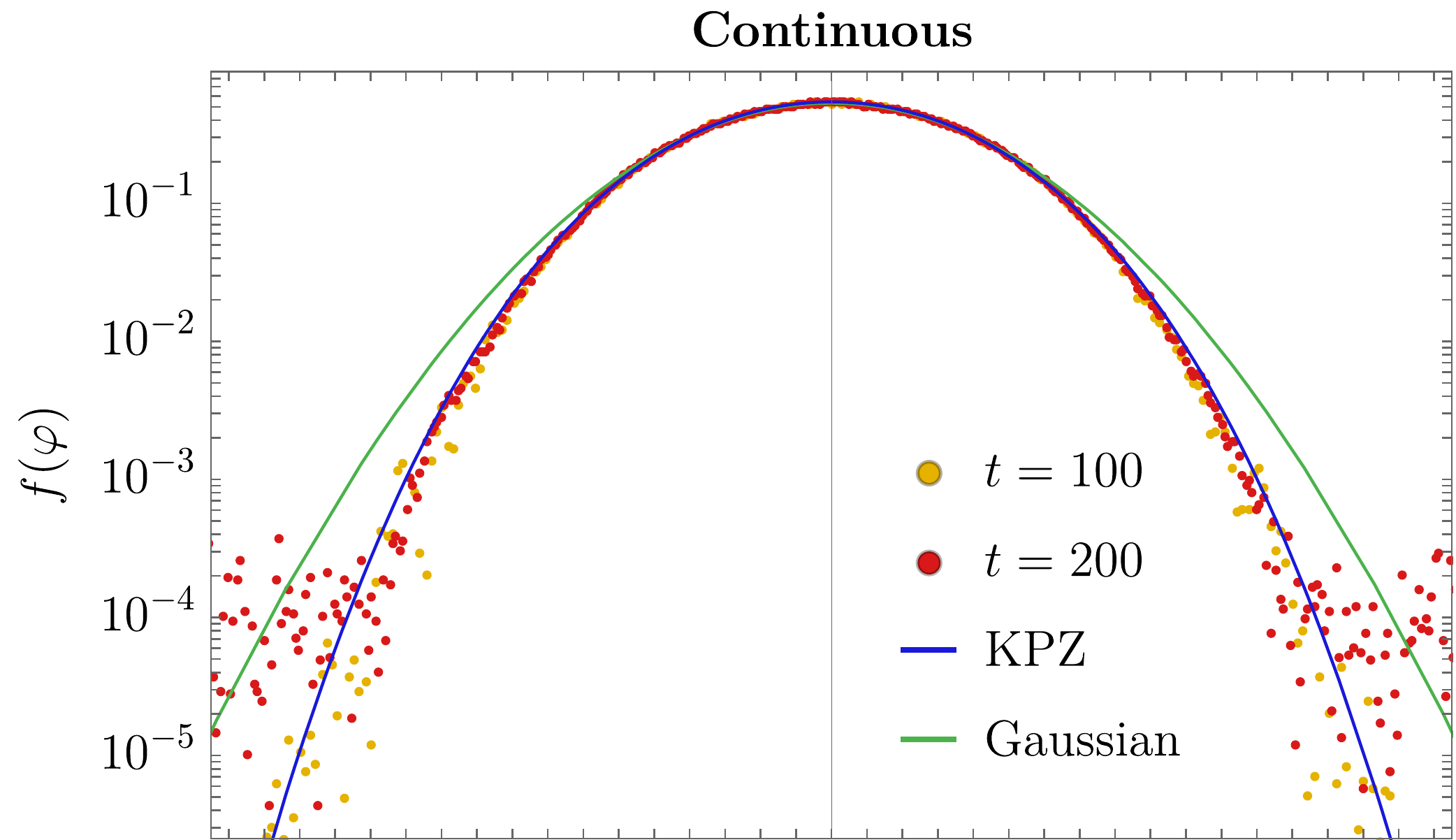}
        \includegraphics[width=0.49\linewidth]{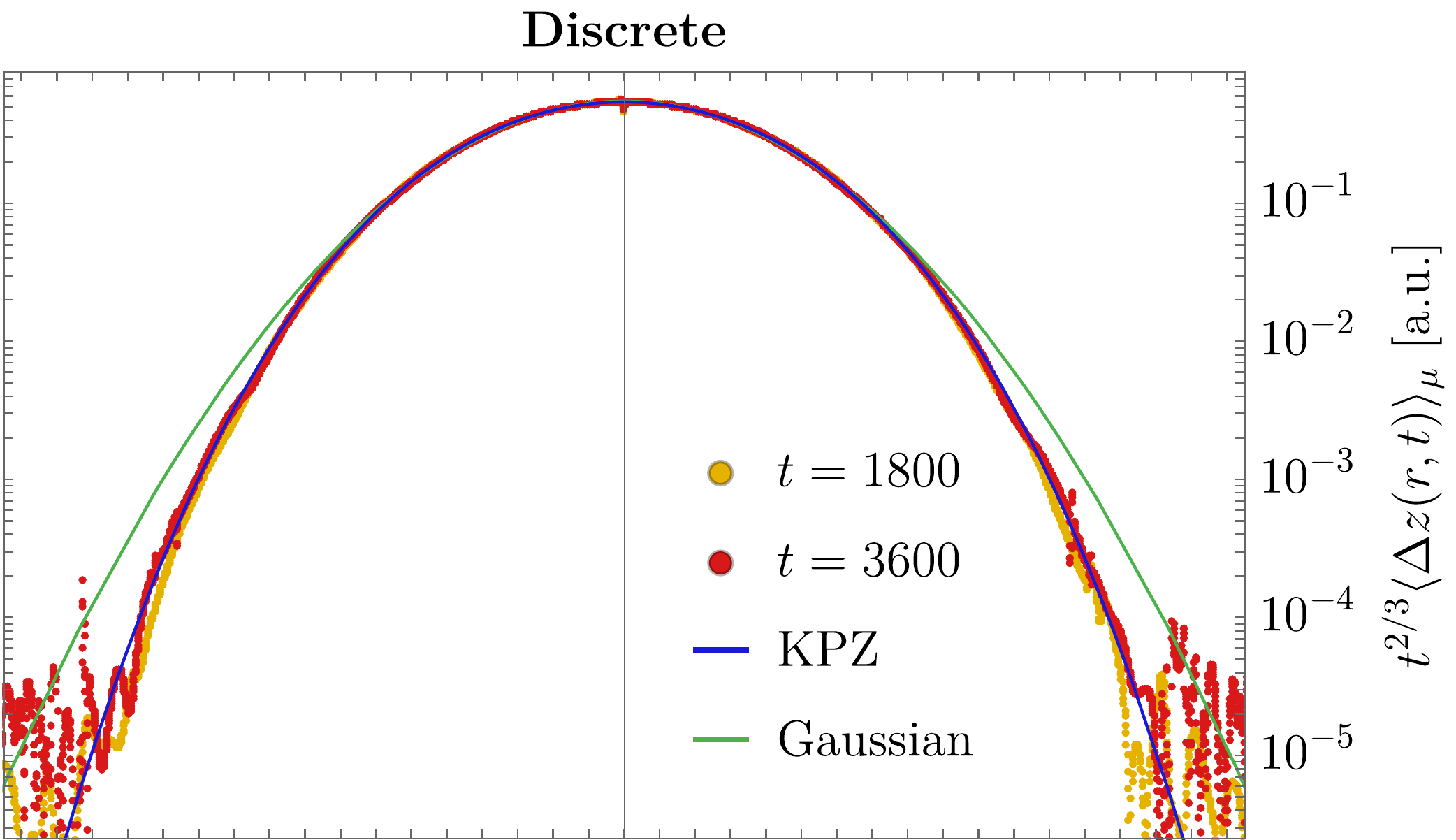}
        \includegraphics[width=0.49\linewidth]{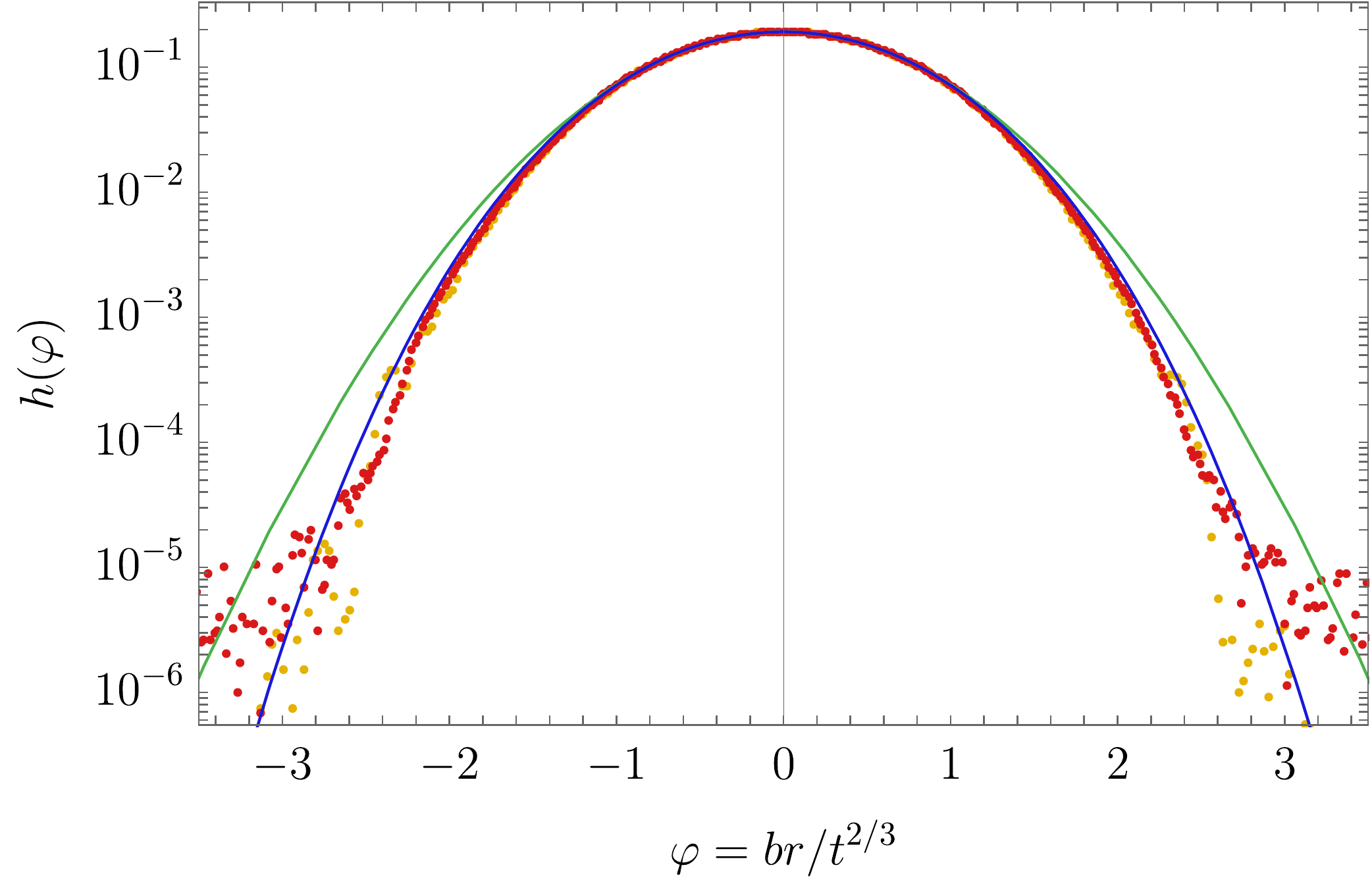}
        \includegraphics[width=0.49\linewidth]{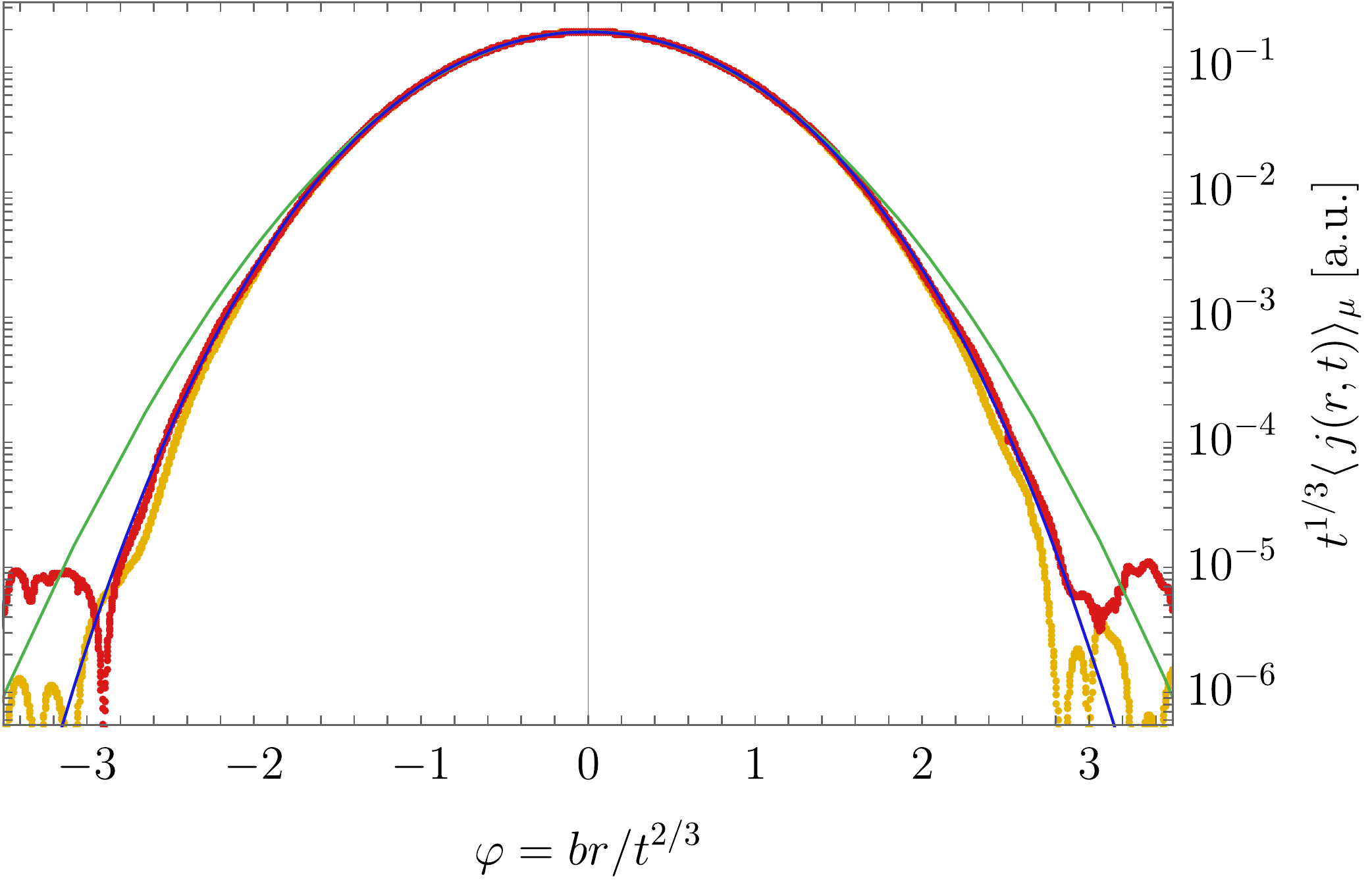}
        \caption{\noindent
            Scaling functions and numerical data: the left column corresponds to the continuous-time model while the right corresponds to the discrete-time model. 
            We show data for the spin current density $\langle\,j\,\rangle_\mu$ and the discrete spin derivative $\Delta z$, defined as $\Delta z=-(\langle\s_r^\z\rangle_\mu-\langle\s_{r-1}^\z\rangle_\mu)$ in the continuous-time model and $\Delta z=-\frac{1}{4}\left(\langle\s_{r+1}^\z\rangle_\mu+\langle\s_{r}^\z\rangle_\mu-\langle\s_{r-1}^\z\rangle_\mu-\langle\s_{r-2}^\z\rangle_\mu\right)$ in the discrete-time model.
            All numerical data (yellow and red points) are appropriately scaled to the KPZ scaling functions, see Eq. (\ref{eq:scl1}) and Eq. (\ref{eq:scl3}). 
            The blue curves represent the KPZ scaling functions while the green ones are the best fitting Gaussian profiles.
            We note that relatively long times are needed in order to observe the KPZ scaling, namely $t\gtrapprox 50$ for the continuous-time model and $t\gtrapprox600$ for the discrete-time model. 
        }
        \label{fig:kpz}
    \end{figure*}
    It shows that a weak domain wall initial state can be seen as a trick that allows us to calculate the infinite-temperature spin-spin correlation. 
    We next recall~\cite{spohnrev} why the LHS of Eq.(\ref{eq:lin3}) is in certain classical systems described by the KPZ scaling function. 
    
    \textbf{Kardar-Parisi-Zhang equation.--} The KPZ stochastic partial differential equation was initially suggested to model the growth of surface $h(r,t)$ through random deposition~\cite{kardar86} 
    \begin{equation}
        \partial_t h=\frac{1}{2}\lambda\left(\partial_rh\right)^2+\nu\partial_r^2h+\sqrt{\Gamma}\zeta\,,
        \label{eq:kpz}
    \end{equation}
    where $\zeta(r,t)$ is a space-time uncorrelated noise. 
    
    Of particular interest to us will be the correlation function $C(r,t)=\langle\left[h(r,t)-h(0,0)-t\langle\partial_th\rangle\right]^2\rangle$ -- representing the fluctuations of the height around the expected value -- and its second derivative $\frac{1}{2}\partial_r^2C(r,t)=\langle\partial_rh(0,0)\partial_rh(r,t)\rangle$ -- describing the slope correlations (here brackets denote noise averaging). In terms of scaling functions $g(\varphi)$  and $f(\varphi)$ one has
    \begin{equation}
        \begin{aligned}
            g(\varphi)&=\lim_{t\to\infty}\frac{C\left((2\lambda^2t^2\Gamma\nu^{-1})^{-1/3}\varphi,t\right)}{\left(\frac{1}{2}\lambda t\Gamma^2\nu^{-2}\right)^{2/3}}\,,\\
            f(\varphi)&=\frac{1}{4}g''(\varphi)\sim\partial_r^2C(r,t)\,.
        \end{aligned}
        \label{eq:kpz_scaling}
    \end{equation}
    These can be obtained from the exact solution of the polynuclear growth model~\cite{spohn04} (a model in the KPZ universality class), and have been tabulated with high precision in Ref.~\cite{prahoferwww}. 
    Nonlinear fluctuating hydrodynamics predicts that the correlation function of a conserved quantity, in our case $\langle\s_0^\z(0)\s_r^\z(t)\rangle$, should be given by the so-called KPZ scaling function $f(\varphi)$. 
    
    Using Eq.(\ref{eq:lin3}) this correlation function is equal to the magnetization difference on consecutive sites in the state $\rho(t) \propto U^t \rho_\mu U^{-t}$ (see Fig.~\ref{fig:profiles}). 
    However, taking the discrete derivative increases numerical errors, so alternatively, one can also look at the scaling form of the current $j(r,t)=\langle\op{j}_r(t)\rangle_\mu$.
    In a diffusive process, the scaling forms of both the current as well as of the magnetization difference are Gaussian. 
    Relation between the two in a general non-diffusive situation can be derived from the continuity equation. 
    
    Defining a shorthand notation $z(r,t)=\langle\s_r^\z(t)\rangle_\mu$, and $\varphi=b\xi$, we write an ansatz
    \begin{equation}
        \partial_rz(r,t)=\frac{a\mu}{t^{2/3}}f\left(\frac{br}{t^{2/3}}\right)\,,
        \label{eq:scl1}
    \end{equation}
    where we introduced two system-dependent parameters $a$ and $b$, and use continuum notation for the magnetization difference. 
    Taking into account the continuity equation $\partial_tz=-\partial_rj$, one may obtain the shape of the spin current profile. 
    Expressing everything in terms of $g(\varphi)$ (using per-partes integration and Eq.~(\ref{eq:kpz_scaling})) we get
    \begin{equation}
        \begin{aligned}
            j(r,t)&=\frac{2a\mu}{3b^2t^{1/3}}h\left(\frac{br}{t^{2/3}}\right)\\
            h(\varphi)&=\frac{g(\varphi)-\varphi g'(\varphi)}{4}\,.
        \end{aligned}
        \label{eq:scl3}
    \end{equation}
    The form of $j(r,t)$, i.e. the function $h(\varphi)$, is therefore uniquely determined by the form of $\partial_r z(r,t)$, i.e., the KPZ function $f(\varphi)$.
    
     \begin{figure}[t!]
        \centering
        \includegraphics[width=0.95\linewidth]{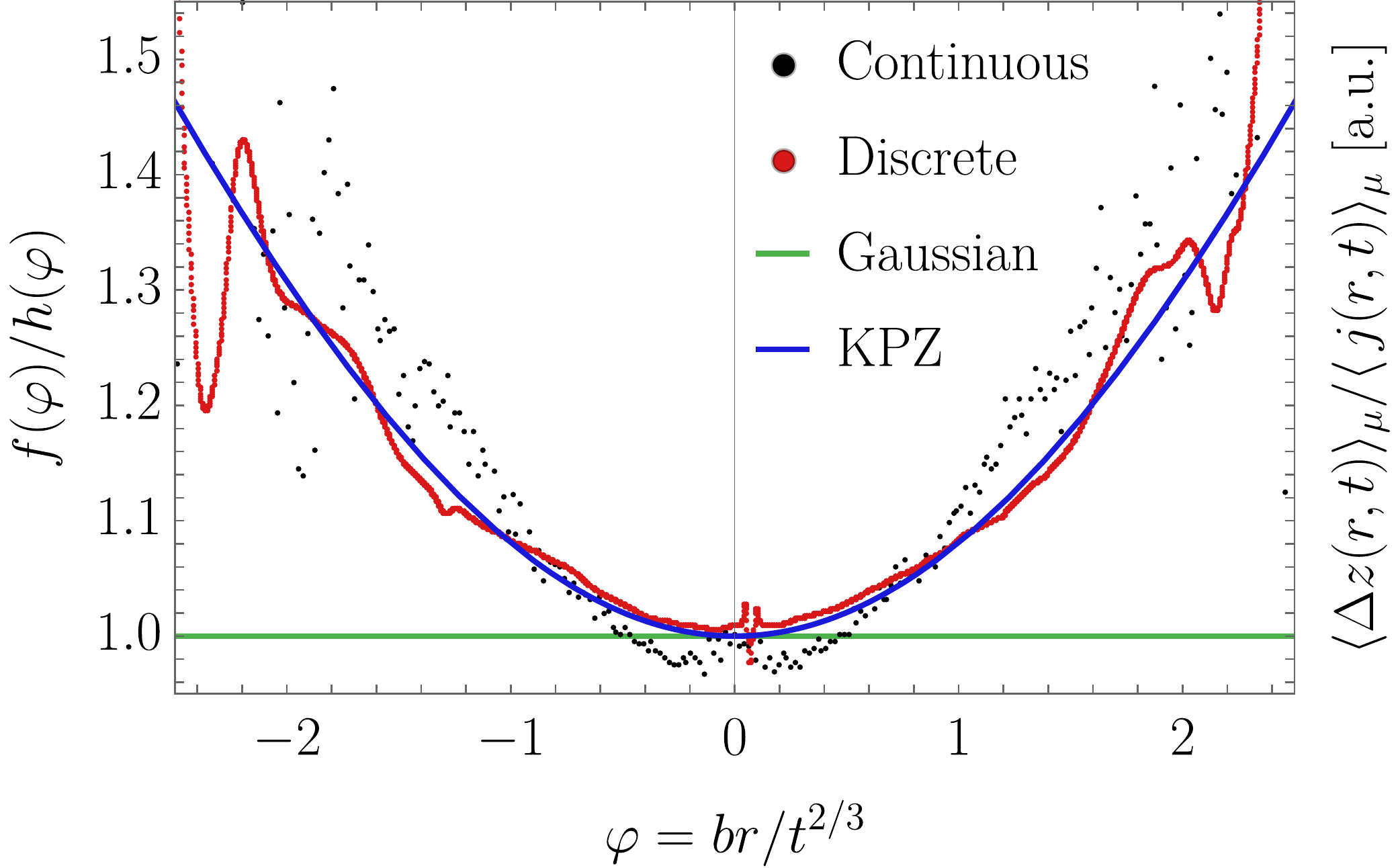}
        \caption{\noindent
            Plotting the ratio between the gradient of spin density and spin current density in scaled units, we can observe that the numerical results for both models clearly do not obey Fick's law. 
            Instead, they are well described by the prediction from KPZ. Numerical data are plotted for maximum simulations times ($t=200$ for continuous and $t=3600$ for discrete time cases). 
            The ratios are rescaled to $1$ at $\varphi=0$. 
        }
        \label{fig:fick}
    \end{figure}
    We employ extensive numerical simulations~\cite{foot3} using the time-evolving block decimation algorithm \cite{vidal03, vidal04, schollwock11} for matrix-product density operator in order to study the time evolution of a domain-wall like initial state in both the continuous and discrete time Heisenberg models. 
    This allows us to compute the infinite-temperature spin-spin correlations (\ref{eq:lin3}) in a numerically stable way with manageable bond dimensions $\chi$. 
    Fig.~\ref{fig:kpz} shows the results and the best-fitting KPZ profile for both the spin and spin current. 
    Due to higher numerical accuracy we only fit the data for the current, obtaining $a$ and $b$ (\ref{eq:scl3}), which automatically fixes the spin difference profiles (\ref{eq:scl1}). 
    In order to avoid even-odd staggering in the discrete-time model we take the difference of two consecutive pairs of spins, rather than a difference of two spins, and appropriately scale the continuity equation. 
    For comparison we also show best-fitting Gaussians. 
    Because the KPZ scaling functions $f(\varphi)$ and $h(\varphi)$ are rather close to Gaussians for not too large arguments, one in fact needs at least two decades of accuracy to be able to distinguish the two. 
    With our numerics we have accuracy over about three decades in the continuous model and about four in the discrete one. 
    We can clearly confirm that the KPZ scaling functions emerge at sufficiently long times.

    Free parameters $a$ and $b$ are found to be $a=b\approx0.67$, conjectured to be $\frac{2}{3 J^{2/3}}$,
    for the continuous-time model.
    Similarly, for the discrete-time model we find $a=b\approx0.43$, data for other values of $J$ are well described by the formula $a=b\approx\frac{2^{1/3}}{3|\tan(J/2)^{2/3}|}$.
    
    Because the KPZ $f(\varphi)$ and $h(\varphi)$ are not Gaussian, their ratio $h/f\equiv w(br/t^{2/3})$ which appears in a relation $j(r,t)=[2t^{1/3}/(3b^2)]w(br/t^{2/3}) \,\partial_r z(r,t)$ is not a constant. Therefore, Fick's law, even with a time-dependent diffusion constant~\cite{foot2}, is violated (Fig.~\ref{fig:fick}).
    \begin{figure}[t!]
        \centering
        \includegraphics[width=0.98\linewidth]{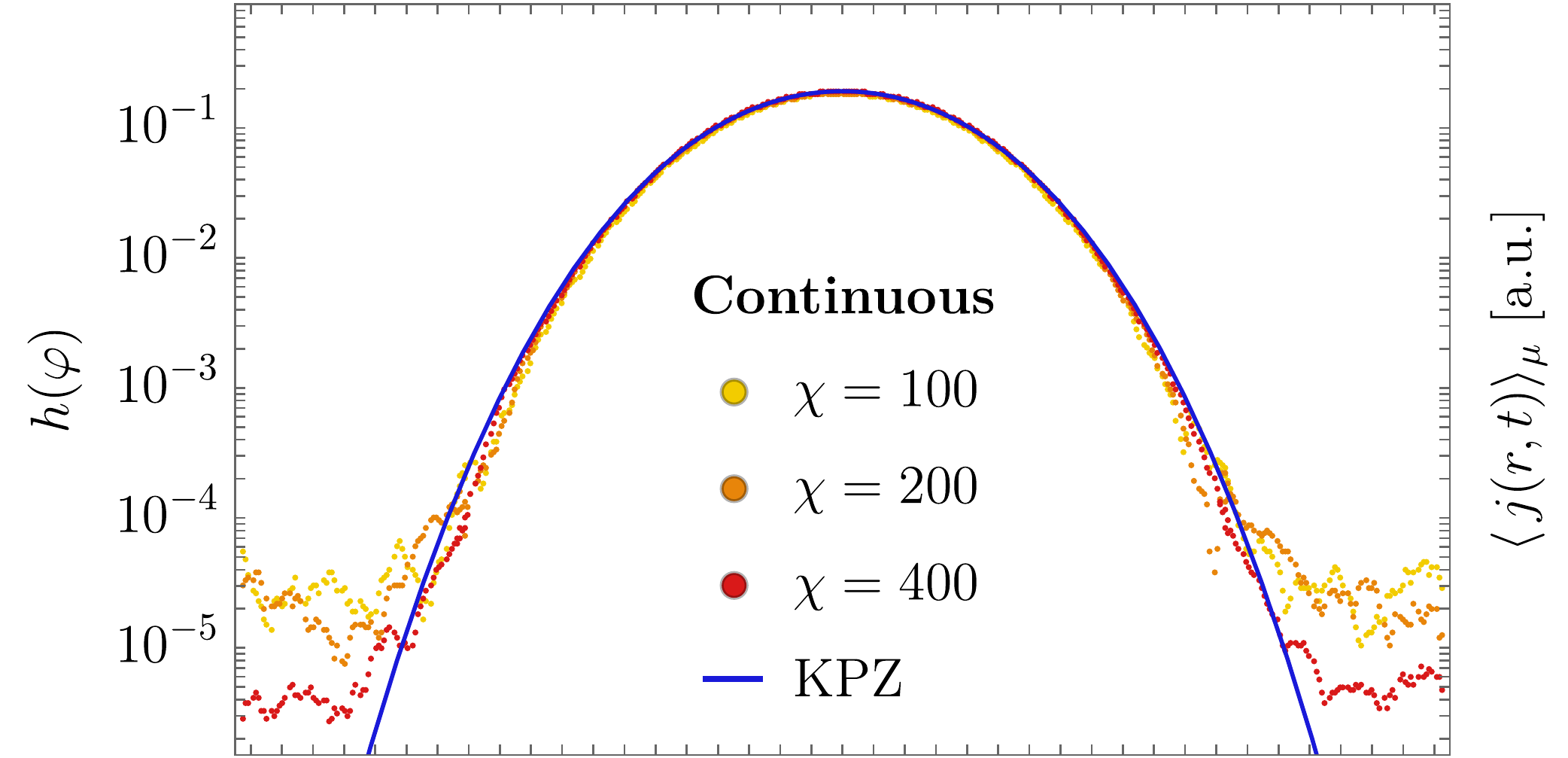}
        \includegraphics[width=0.98\linewidth]{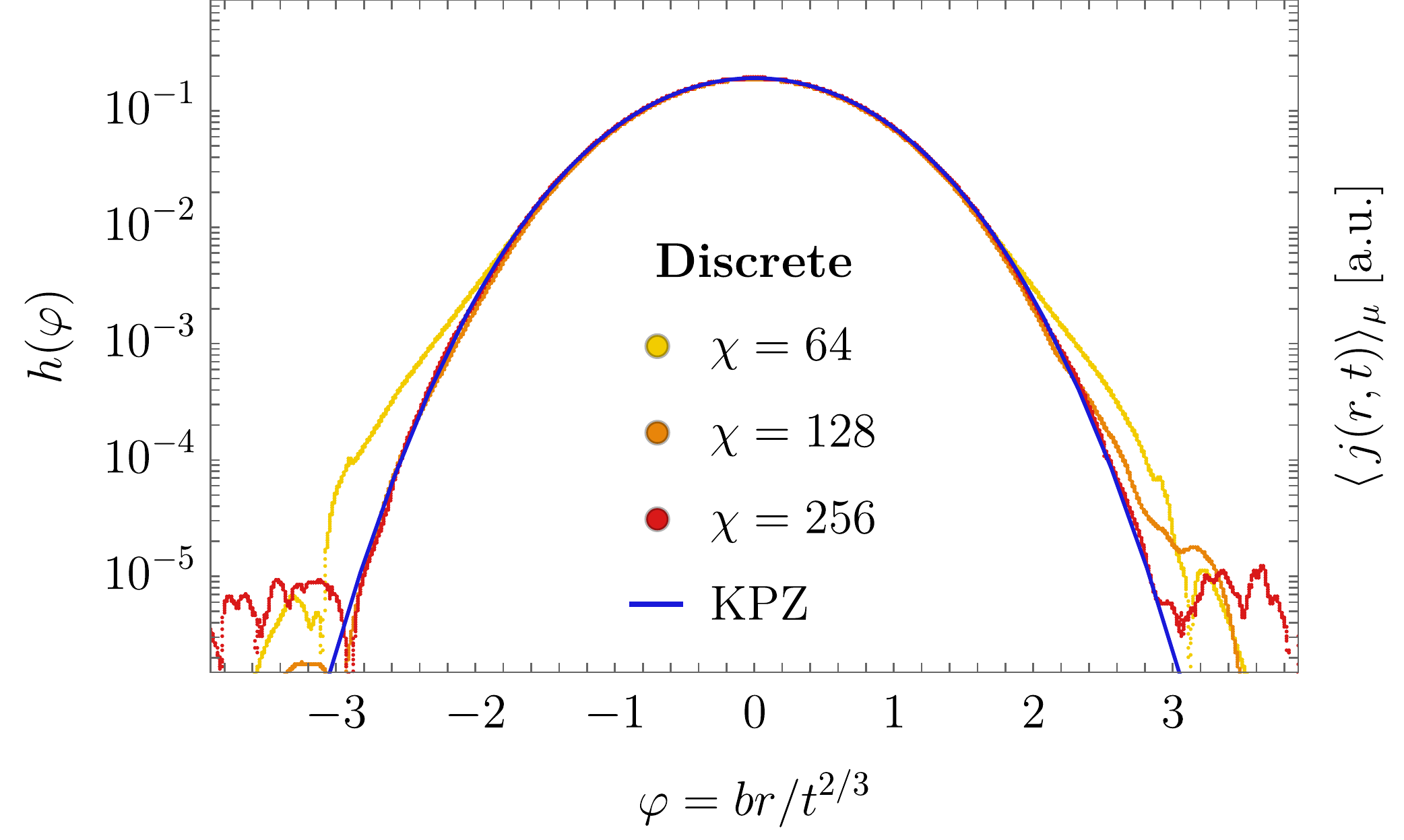}
        \caption{\noindent
            Dependence on bond dimension of the current profiles in a domain-wall state and for continuous ($t=200$) and discrete-time simulations ($t=3600$). 
            Results are stable to increasing $\chi$ and converge to the KPZ scaling functions. 
            We apply a moving average to the leftmost and rightmost $20\%$ of the data so that it is easier to see the decreasing truncation error in the tails.
        }
        \label{fig:bond}
    \end{figure}
    
    We also show the dependence of current profiles on the bond dimension $\chi$ used in simulations, Fig.~\ref{fig:bond}. 
    In the discrete-time case we use slightly smaller $\chi$, however the acquired times     are larger (Fig.~\ref{fig:kpz}), as well as the sizes ($L=7200$ vs. $L=400$). 
    As a net result the wall-times of discrete model simulations are about half as long as for a continuous one despite about a decade better accuracy (Fig.~\ref{fig:bond}). 
    We stress that in the best classical simulations (hard-point gas~\cite{mendl14}) slightly less than two decades of agreement with KPZ are achieved. 
    What distinguishes our quantum simulations is that we directly work with an ensemble, encoded in the many-body density matrix $\rho(t)$, so no averaging is needed. 
    It is an interesting open problem how to do such efficient ensemble simulations for classical many-body models, in particular since for continuous variables the local function spaces are infinitely dimensional.

    Lastly, we note that taking a slightly larger domain-wall step $\mu=0.02$ we are even able to observe (data not shown) second order $\mu^2$ corrections to the dynamics in the form of a small ballistically spreading front, traveling away from the site of the quench. 
     
    \textbf{Discussion.--}
    We have shown that the infinite temperature spin-spin correlation function in the isotropic Heisenberg spin-$\frac{1}{2}$ model obeys the Kardar-Parisi-Zhang scaling. 
    This is the first such observation in a deterministic quantum model. 
    We stress that in order to reliably show the KPZ physics one has to look at the full distribution function of fluctuations and not e.g. just the dynamical scaling exponent being $z=\frac{3}{2}$. 
    For instance, a related spreading exponent $\frac{1}{3}$ generically appears in free or dilute models, see e.g.~\cite{racz04,vir18}.
    
    High accuracy of over four decades was achieved by using a trick where we simulate the melting of a slightly polarized domain wall by directly evolving the density operator, which is, through linear response, equivalent to studying the equilibrium spin-spin correlation function.
        
    Besides providing a method to efficiently probe spatio-temporal correlation functions in quantum models, several new directions are opened. 
    The most important is the question of universality. 
    Namely, in nonlinear fluctuating hydrodynamics, so-far verified only in classical models, the KPZ universality is associated to the existence of 3 conservation laws. 
    It is not clear which 3 conserved quantities (if at all) are responsible for the observed behavior. 
    By studying a kicked Floquet generalization of the isotropic Heisenberg model which does not conserve the energy, but nevertheless shows the KPZ physics, we show that the energy is not one of them. 
    It remains to be seen if the observed behavior is in any way related to integrability and the SU(2) symmetry of the model.

    \textbf{Acknowledgements.--}
    We acknowledge useful related discussions with J. De Nardis, E. Ilievski, M. Medenjak, and H. Spohn. 
    The authors acknowledge support by the European Research Council (ERC) through the advanced grant 694544 – OMNES and the grants P1-0402 and J1-7279 of the Slovenian Research Agency (ARRS).


\begin{thebibliography}{99}
\bibitem{foot2} In Ref.~\cite{ljubotina17} our accuracy was less than two decades due to smaller bond dimensions $\chi$ used, hence we could not reliably discriminate between KPZ and Gaussian scaling functions.
\end{thebibliography}
\end{document}